\documentclass[pre,aps,amsmath,amssymb,floatfix,twocolumn,nofootinbib]{revtex4-1} 

\usepackage{ifthen} 
\newboolean{retro} 
\setboolean{retro}{true} 

\ifthenelse{\boolean{retro}}{ 

\usepackage{background}
\usepackage{cmbright}
\usepackage{caption} 
\usepackage{courier}
\usepackage{titlesec}
\setlength{\fboxrule}{0.1pt}

\usepackage{geometry}
\geometry{
  left=2cm,
  right=2cm,
  top=2.5cm,
  bottom=2.5cm
}

\usepackage{setspace}
\setstretch{1.2} 

\usepackage{fancyhdr}
\pagestyle{fancy}
\fancyhf{}
\rhead{\thepage}
\lhead{Nonequilibrium erasure at varying temperature}

\backgroundsetup{scale=1,opacity=0.25,angle=0,contents={\includegraphics[width=\paperwidth,height=\paperheight]{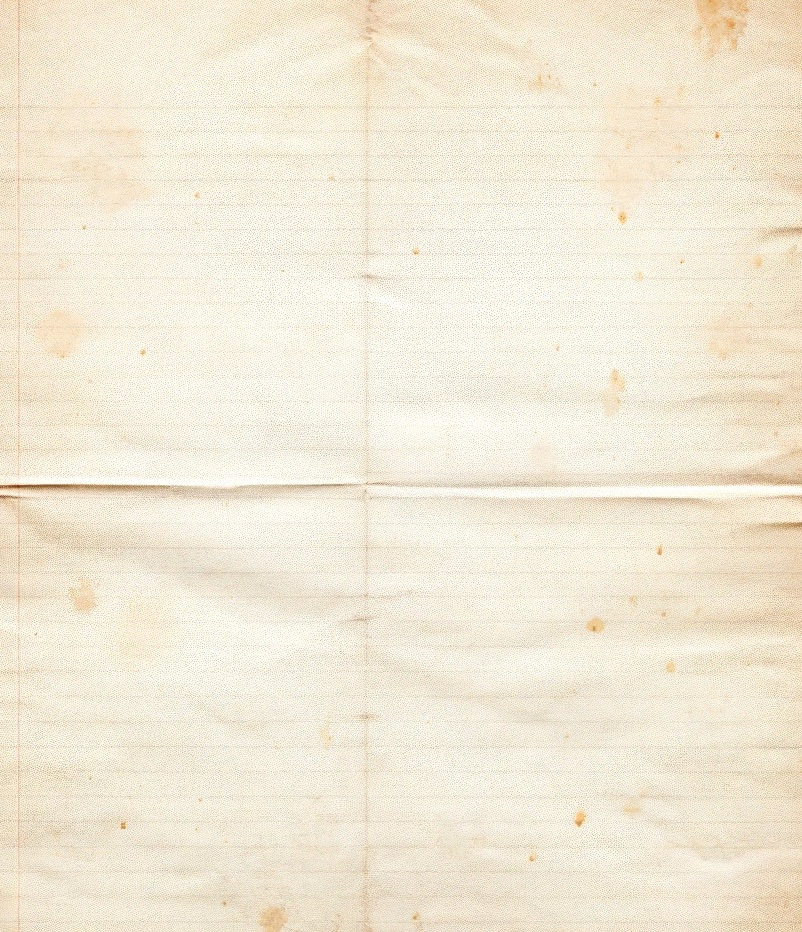}}}

\captionsetup[figure]{font={tt,small}}

\titleformat{\section}{\normalfont\normalsize}{\thesection}{1em}{}
\titleformat{\subsection}{\normalfont\normalsize}{\thesubsection}{1em}{}
\titleformat{\subsubsection}{\normalfont\normalsize}{\thesubsubsection}{1em}{}
}{
\usepackage[usenames]{color}
\usepackage{hyperref}
}

\pdfoutput=1
\usepackage{bm}
\usepackage{graphicx}

\def\eqq#1{Eq.~(\ref{#1})}

\def\eq#1{(\ref{#1})}

\def\f#1{Fig.~\ref{#1}}

\def\s#1{Section~\ref{#1}}

\def\c#1{~\cite{#1}}

\def\cc#1{Ref.~\cite{#1}}

\def\av#1{\langle #1 \rangle}

\def\beq{\begin{equation}}
\def\eeq{\end{equation}}
\def\bea{\begin{eqnarray}}
\def\eea{\end{eqnarray}}

\def\kt{k_{\rm B}T}
\def\cee{{\bm c}}
\def\tf{t_{\rm f}}
\def\e{{\rm e}}

\begin{document}

\title{\normalfont Nonequilibrium formulation of varying-temperature bit erasure}
\author{Stephen Whitelam}
\affiliation{Molecular Foundry, Lawrence Berkeley National Laboratory, 1 Cyclotron Road, Berkeley, CA 94720, USA}

\begin{abstract}
Landauer's principle states that erasing a bit of information at fixed temperature $T$ costs at least $\kt \ln 2$ units of work. Here we investigate erasure at varying temperature, to which Landauer's result does not apply. We formulate bit erasure as a stochastic nonequilibrium process involving a compression of configuration space, with physical and logical states associated in a symmetric way. Erasure starts and ends at temperature $T$, but temperature can otherwise vary with time in an arbitrary way. Defined in this way, erasure is governed by a set of nonequilibrium fluctuation relations that show that varying-temperature erasure can done with less work than $\kt \ln 2$. As a result, erasure and the complementary process of bit randomization can be combined to form a work-producing engine cycle.
\end{abstract}
\maketitle

\section{Introduction}
Landauer's principle states that resetting a one-bit memory at temperature $T$ results in the emission of at least $\kt \ln 2$ units of heat, and costs at least $\kt \ln 2$ units of work\c{landauer1961irreversibility,sagawa2014thermodynamic}. This principle has been verified experimentally, and places limits on the efficiency of irreversible computation\c{bennett1982thermodynamics,bennett1985fundamental,piechocinska2000information,dillenschneider2009memory,jun2014high,berut2012experimental,dago2021information,hong2016experimental,frank2018physical,frank2005approaching,still2012thermodynamics,maroney2005absence,maroney2009generalizing,seet2023simulation}. 

However, Landauer's bound does not apply if temperature varies\c{wolpert2023stochastic}, which theory and simulation suggests is a characteristic of optimal bit-erasure protocols\c{gingrich2016near,rotskoff2015optimal,engel2023optimal}. In this paper we investigate bit erasure at varying temperature. We use a Langevin particle in an external potential as a model of a one-bit memory, by associating physical states with logical states in a symmetric way. We define erasure in a fluctuating nonequilibrium setting, as a finite-time transformation from an initial double-well potential, in which the particle starts in thermal equilibrium, to a final single-well potential.

Defined in this way, erasure at fixed temperature is governed by the Crooks\c{crooks1999entropy} and Jarzynski\c{jarzynski1997nonequilibrium} nonequilibrium fluctuation relations\c{das2014capturing}. These relations allow us to calculate the efficiency of a cycle composed of bit erasure and its time-reverse, bit randomization, and to show that the mean work required to achieve erasure is bounded by a quantity that approaches $\kt \ln 2$ as the potential barrier becomes large. This bound is a statement of the second law of thermodynamics, and is the analog of the Landauer bound for the nonequilibrium formulation of erasure that we consider.

We then allow temperature to vary during erasure, with the starting and ending temperatures equal to the value $T$ used in the fixed-temperature erasure scheme. We show that erasure at varying temperature obeys the varying-temperature fluctuation relations of~\cc{whitelam2023free}. We use these relations to show that the work required to do erasure is not bounded by $\kt \ln 2$, and that erasure and randomization can operate as a work-producing engine.

In \s{model} we introduce the model memory used in this paper, a Langevin particle in an external potential. In \s{const} we consider erasure at fixed temperature, to which standard fluctuation relations apply. In \s{final} we comment on the effect of ending erasure with a double-well form of the potential, which is standard, rather than the single-well form of the potential used in this paper. In \s{vary} we carry out erasure at varying temperature, and analyze it using nonequilibrium fluctuation relations valid for varying-temperature protocols. We conclude in \s{conc}.

\section{Model and simulation details}
\label{model}

We consider a particle at position $x$ undergoing the Langevin dynamics
\beq
\label{langevin}
\dot{x}=-\partial_x U(\cee,x) + \xi(t),
\eeq
where $\av{\xi(t)}=0$ and $\av{\xi(t) \xi(t')} = 2 \beta(t)^{-1} \delta(t-t')$. The temperature of the system is therefore $T(t)=\beta(t)^{-1}$, which can vary with time (henceforth we work in units in which $k_{\rm B}=1$). The potential $U(\cee,x)$ is
\beq
\label{pot}
U(\cee,x) = \frac{k}{2} \left( |x-c_0|-c_1\right)^2,
\eeq
parameterized by the coefficients $\cee=(c_0,c_1)$ and the spring constant $k$. For $c_1\neq 0$ this potential has a double-well form with minima at $x=c_0 \pm c_1$ and barrier height $k c_1^2/2$. For $c_1=0$ the potential has a single-well form with its minimum at $x=c_0$. Unless otherwise stated we set $k=20$. We can consider the model to define a 1-bit memory if we associate positions $x \leq 0$ and $x> 0$ with logical states $s=0$ and 1, respectively\footnote{The association of the case $x=0$ with state $s=0$ but not state $s=1$ means that we associate positions and logical states in a (very slightly) asymmetric manner. But since $x$ is real-valued, this asymmetry has no practical significance.}.

 The initial reciprocal temperature of the system is $\beta(0)=1$. The initial potential has coefficients $\cee(0)=(0,1)$, which has the double-well form shown in panel A of \f{fig1}(b). The particle starts in equilibrium with this potential, and so the distribution of its initial positions is that of the Boltzmann distribution $\rho_0(\cee(0),x)$, where $\rho_0(\cee(t),x)=Z^{-1}(\cee(t)) \, \e^{-\beta(t) U(\cee,x)}$ and $Z(\cee(t)) =\int {\rm d} x' \,\e^{-\beta(t) U(\cee,x')}$.
\begin{figure*}[] 
   \centering
\ifthenelse{\boolean{retro}}{\fbox{\includegraphics[width=\linewidth]{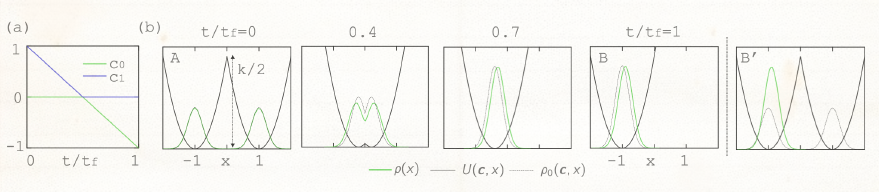}}}
     {\includegraphics[width=\linewidth]{fig1}}
       \caption{(a) Erasure protocol \eq{def1}. (b) Effect of the erasure protocol for trajectory time $\tf=1$, showing potential (black), associated Boltzmann distribution (black dashed), and instantaneous particle-position distribution (green), calculated over $10^6$ trajectories. Erasure ends with the form of the potential labeled B; we also discuss the effect of switching back to the double-well form B$'$.}
   \label{fig1}
\end{figure*}

Erasure trajectories are run for time $\tf$ (note that the characteristic time for a free particle to diffuse a distance equal to the separation between potential wells is $t_0=2$ for $\beta(t)=1$). We consider a memory-reset protocol $\cee(t)$ similar to that used in Refs.\c{dago2021information,dago2022virtual}, namely
\bea
\label{def1}
\cee(t)= \begin{cases}
(0,1-2 t/\tf) & 0\leq t/\tf<1/2 \\
(1-2t/\tf,0) & 1/2 \leq t/\tf \leq1,
\end{cases}
\eea
which is plotted in \f{fig1}(a). As shown in \f{fig1}(b), this protocol merges the potential wells and slides the resulting single well to the left, until its minimum coincides with the left-hand minimum of the double-well form of the potential. At time $t=\tf$ the potential has the coefficients $\cee(\tf)=(-1,0)$ and the single-well form shown in panel B of \f{fig1}(b). (We shall also consider the effect of switching back to the initial form of the potential, as shown in panel B$'$ of \f{fig1}(b).) At time $\tf$, temperature is returned to its initial value, $\beta(\tf)=1$. 

The free-energy change $\Delta F_{\rm e}$ for the transformation from double- to single-well forms during erasure (A$\to$B in \f{fig1}(b)) is $\Delta F_{\rm e} = -\beta^{-1} \ln [Z(\cee(\tf))/Z(\cee(0))]$, giving
\beq
\label{fec}
\beta \Delta F_{\rm e} = \ln \left( 2-{\rm erfc} \sqrt{\beta k/2}\right),
\eeq
where ${\rm erfc} \, x \equiv 2 \pi^{-1/2} \int_x^\infty {\rm d}t \, \e^{-t^2}$ is the complementary error function. For $\beta=1$ (the start and end temperature of our erasure protocol) and large spring constant $k$, we have 
\beq
\Delta F_{\rm e} \approx \ln 2 -(2 \pi k)^{-1/2}\e^{-k/2},
\eeq
which approaches $\ln 2$ as $k$ diverges. For the value $k=20$, $\Delta F_{\rm e}$ is less than $\ln 2$ by about $4 \times 10^{-6}$.

 We shall also consider the time-reverse of \eq{def1}, $\cee_{\rm r}(t) = \cee(\tf-t)$. This protocol effects randomization of a bit set initially to state 0, starting in equilibrium with the potential in the single-well form with $\cee_{\rm r}(0)=(-1,0)$, shown in panel B of \f{fig1}(b). The free-energy change for the randomization protocol is the negative of \eq{fec}.

We integrate \eqq{langevin} using a first-order Euler scheme with timestep $\Delta t=10^{-3}$\ifthenelse{\boolean{retro}}{, evaluated by electronic computing machine}{}. At step $i=1,2,\dots,N=\lfloor \tf/\Delta t \rfloor$ of the simulation the time is $t_i=i \Delta t$, the position of the particle is $x_i$, and the values of the coefficients are $\cee_i$. The work done in a single trajectory is 
\beq
W= \sum_{i=0}^{N-1} [U(\cee_{i+1},x_i)-U(\cee_i,x_i)],
 \eeq
and we shall consider averages $\av{W}$ and distributions $P(W)$ taken over $10^6$ independent trajectories. The probability of erasure is $p_{\rm e} = \av{\Theta(-x_N)}$, where $\Theta(-x)=1$ if $x \leq 0$ and $\Theta(-x)=0$ otherwise.

\section{Constant-temperature protocols}
\label{const}

We first consider constant-temperature erasure and randomization protocols for which $\beta(t)=1$ for all $t$. \f{fig2} shows the results of simulations carried out for a range of trajectory times $\tf$; averages and distributions are calculated over $10^6$ independent trajectories.
\begin{figure*}[] 
   \centering
        \ifthenelse{\boolean{retro}}{\fbox{\includegraphics[width=\linewidth]{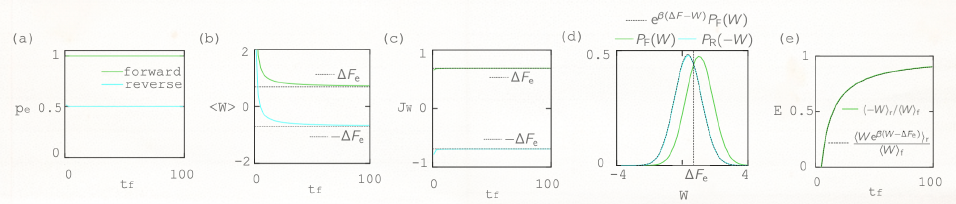}}}
          {\includegraphics[width=\linewidth]{fig2}}
  \caption{Constant-temperature erasure protocol shown in \f{fig1} (green) and its time reverse, randomization (cyan), for various trajectory times $\tf$. (a) Probability of erasure. (b) Mean work done, which satisfies the second law of thermodynamics, \eqq{av}. (c) Jarzynski free-energy estimator \eq{est}. (d) Crooks fluctuation relation \eq{fluc} for trajectory length $\tf=10$. (e) Efficiency of the forward-reverse erasure procedure, calculated using both forward and reverse protocols (green) and the forward protocol only (black dashed). The color scheme in (e) is distinct from that in the other panels.}
   \label{fig2}
\end{figure*}

In \f{fig1}(a) we show the probability of erasure (the probability that the particle has coordinate $x<0$ at time $\tf$), which is close to 1 and $1/2$ for the erasure and randomization protocols, respectively, For the erasure protocol, 1  or 2 in a million trajectories results in a failed erasure (we will discuss the significance of this fact more in the following section). In panel (b) we show the mean work $\av{W}$ under the two protocols, which obeys the second law of thermodynamics in the form\c{sagawa2014thermodynamic}
\beq
\label{av}
\av{W} \geq \Delta F.
\eeq
Here $\Delta F$ is the free-energy change associated with the protocol, which is $\Delta F_{\rm e}$ for erasure and $-\Delta F_{\rm e}$ for randomization. The work values in panel (b) approach the asymptotes $\Delta F_{\rm e}$ and $-\Delta F_{\rm e}$ as $\sim 1/\tf$.

In \f{fig1}(c) we show that the protocols obey the Jarzynski equality\c{jarzynski1997nonequilibrium}
\beq
\label{jarz}
\av{\e^{-\beta W}} = \e^{-\beta \Delta F},
\eeq
by plotting the free-energy estimator
\beq
\label{est}
J_W = -\beta^{-1} \ln \left( N_{\rm traj}^{-1} \sum_{i=1}^{N_{\rm traj}} \e^{-\beta W_i} \right).
\eeq
(Recall that $\beta=1$ for the fixed-temperature erasure protocol.) In \eq{jarz}, the angle brackets denote an average over trajectories that start in thermal equilibrium and enact the specified protocol (with $\Delta F=\pm \Delta F_{\rm e}$ for erasure and randomization, respectively). In \eq{est}, $N_{\rm traj} = 10^6$ is the number of trajectories used, and $W_i$ is the work value associated with trajectory $i$. \f{fig2}(c) shows that \eq{est} provides a good estimate of the free-energy change for erasure and randomization, for all but the smallest value of $\tf$.

In \f{fig1}(d) we show, for one value of $\tf$, that erasure and randomization protocols obey the Crooks work fluctuation relation\c{crooks1999entropy}
\beq
\label{fluc}
P_{\rm F}(W) \e^{-\beta W} = \e^{-\beta \Delta F} P_{\rm R}(-W).
\eeq
Here $P_{\rm F}(W)$ is the probability distribution of $W$ for erasure (``forward'') protocols, while $P_{\rm R}(-W)$ is the probability distribution of $-W$ for randomization (``reverse'') protocols. The quantity $\Delta F=\Delta F_{\rm e}$ refers to the free-energy change for the forward protocol. 

Panels (b), (c), and (d) of \f{fig1} make increasingly detailed statements about work fluctuations for these protocols, illustrating Equations~(\ref{av}),~(\ref{jarz}), and~(\ref{fluc}), respectively.  \eqq{jarz} can be obtained from \eqq{fluc} by integration, while \eqq{av} can be obtained from \eqq{jarz} by application of the Jensen inequality. 

In \f{fig1}(e) we show the efficiency $\av{-W}_{\rm r}/\av{W}_{\rm f}$ of the forward-reverse cycle, the ratio of the work extracted by randomization to that expended during erasure. This quantity is positive for trajectory lengths $\tf \gtrsim 1$, and approaches 1 for large $\tf$. In that limit, a logically irreversible procedure, the resetting and subsequent randomization of a bit, is done in a thermodynamically reversible way\c{maroney2005absence,sagawa2014thermodynamic}. The forward-reverse cycle could be also be considered a battery, with work stored during erasure and extracted during randomization.

Using \eq{fluc} we can rewrite averages over the reverse process in terms of averages over the forward process, $\av{(\cdot)}_{\rm r}=\av{(\cdot)\e^{\beta (W-\Delta F)}}_{\rm f}$, and so compute the efficiency of the erasure-randomization cycle by doing simulations of the forward process only. This fact is shown by the coincidence of the green and black lines in \f{fig1}(e).
 
 \section{Consequences of the final form of the potential}
 \label{final}
 
We have defined the erasure process as one that starts with the double-well potential given by \eq{pot} with coefficients $\cee(0)=(0,1)$,
\beq
U_{\rm d}(x)=U(\cee(0),x)=\frac{k}{2} (|x|-1)^2,
\eeq
and ends with the single-well potential with coefficients $\cee(\tf)=(-1,0)$,
\beq
\label{sw}
U_{\rm s}(x)=U(\cee(\tf),x)=\frac{k}{2} (x+1)^2.
\eeq
The free-energy change $\Delta F_{\rm e}$ at temperature $\beta^{-1}$ associated with this change is given by \eqq{fec}, which is very close to $\beta^{-1} \ln 2$ for the value $k=20$ used in our simulations. The mean work required to effect this change is therefore constrained by the second law, \eqq{av}, and is $\av{W} \geq \Delta F_{\rm e}$.

\begin{figure*}[] 
   \centering
       \ifthenelse{\boolean{retro}}{ \fbox{\includegraphics[width=0.7\linewidth]{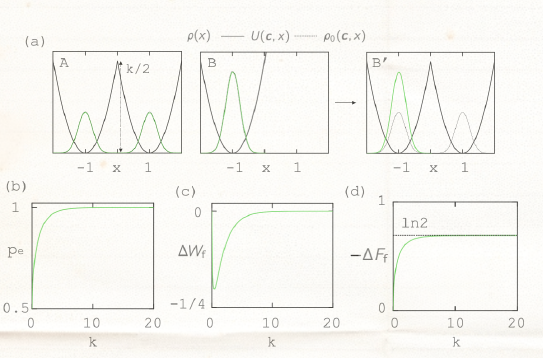}}}
       {\includegraphics[width=0.7\linewidth]{fig_s1}}
  \caption{Effect of switching back to a double-well potential after the erasure protocol of \f{fig1}. (a) Slow transformation from A$\to$B, followed by a sudden switch to B$'$. (b) Probability of erasure if the final-time particle-position distribution is in equilibrium with the single-well potential B. (c) Change of work upon switching to the double-well form B$'$. (d) Change of free energy (at $\beta=1$) upon switching from B to the double-well form B$'$.}
   \label{fig3}
\end{figure*}

However, is conventional in the literature to formulate erasure problems that start and end with a double-well potential\c{landauer1961irreversibility,sagawa2014thermodynamic,berut2012experimental,dago2021information}. The practical consequences of switching back to the double-well form at the end of the erasure protocol are relatively minor, as long as the trap spring constant $k$ is large enough, but doing so introduces a conceptual complication: the free-energy change is now zero, and so the work required to do erasure is bounded only by zero, $\av{W} \geq 0$.

To illustrate this point, imagine that we carry out the erasure protocol of Figs.~\ref{fig1} and~\ref{fig2} very slowly, so that the particle-position distribution remains in quasiequilibrium with the potential (similar considerations apply to nonequilibrium protocols that result in erasure with probability close to unity). This scenario is sketched in \f{fig3}(a), with panels A and B representing the start and end of the process. Following the transformation A$\to$B, the particle-position distribution for $\beta=1$ is 
\beq
\rho_{\rm s}(x)=\sqrt{\frac{k}{2 \pi}} \e^{-\frac{k}{2} (x+1)^2},
\eeq
reflecting equilibrium with the single-well form of the potential, \eqq{sw}. The probability of erasure is the probability that the particle resides in the sector $x \leq 0$, which is
\beq
\label{pe}
p_{\rm e} = \int_{-\infty}^{0} {\rm d} x\,\rho_{\rm s}(x)=1-\frac{1}{2} {\rm erfc}(\sqrt{k/2}).
\eeq
\eqq{pe} is plotted as a function of trap spring constant $k$ in \f{fig3}(b). The erasure probability $p_{\rm e}$ is less then unity by about $4 \times 10^{-6}$ at $k=20$.
\begin{figure*}[] 
   \centering
      \ifthenelse{\boolean{retro}}{  \fbox{\includegraphics[width=\linewidth]{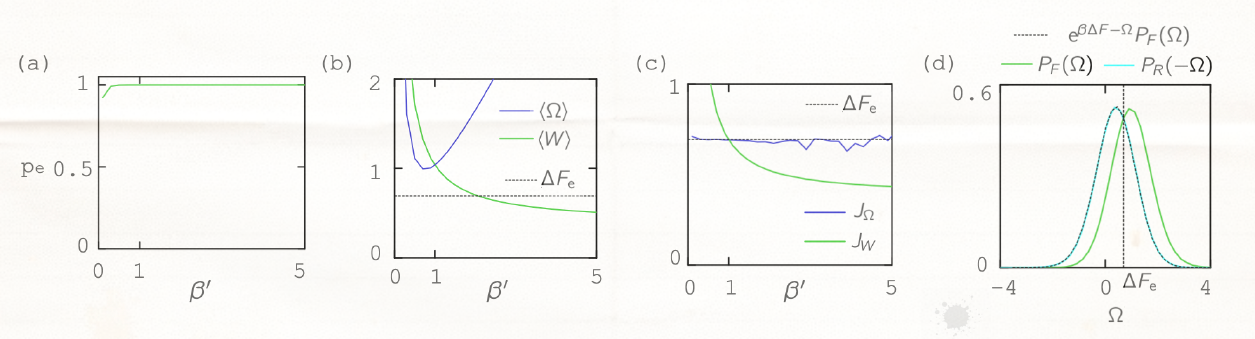}}}
      {\includegraphics[width=\linewidth]{fig3}}
 \caption{Erasure process of \f{fig1}(a) for trajectory time $\tf=10$, now at varying temperature: $\beta(t) = \beta'$ for $0<t<\tf$, with $\beta(0)=\beta(\tf)=1$. (a) Probability of erasure as a function of $\beta'$. (b) The mean of $\Omega$ is bounded by $\Delta F_{\rm e}$, by \eqq{av_om}, but the mean work $\av{W}$ is not. (c) The integral fluctuation relation \eq{jarz_mod} and (d) the fluctuation relation \eq{fluc_mod} hold at varying temperature.}
   \label{fig4}
\end{figure*}

The mean work required to do this erasure is $\av{W} = \Delta F_{\rm e}$ (if done infinitely slowly). We then switch suddenly to the double-well form of the potential, as shown in panel B$'$. What happens?
 
The difference in energy upon changing the single-well potential $U_{\rm s}(x)$ to the double-well potential $U_{\rm d}(x)$ is
\beq
\Delta U(x) \equiv U_{\rm d}(x)-U_{\rm s}(x)=\begin{cases} 0 & (x<0)\\ -2 k x & (x \geq 0) \end{cases},
\eeq
and so the mean change of work caused by the final-time switch in potential is
\bea
\Delta W_{\rm f}&=& \int_{-\infty}^{\infty} {\rm d} x\, \Delta U(x) \rho_{\rm s}(x) = -2 k \int_{0}^{\infty} {\rm d} x\,  x \rho_{\rm s}(x) \nonumber \\
&=& k \,{\rm erfc}(\sqrt{k/2})-\sqrt{\frac{2k}{\pi}} \e^{-k/2}.
\eea
This quantity is plotted as a function of $k$ in \f{fig3}(c). For $k=20$, the mean change of work $\Delta W_{\rm f}$ is negligible, and the work required to perform the process A$\to$B$\to$B$'$ is essentially the same as that required to perform the process A$\to$B, namely $\Delta F_{\rm e}$.

However, the change of free energy upon making the change B$\to$B$'$ is significant. For temperature $\beta^{-1}=1$, the free-energy change resulting from the final change of potential is, by the Zwanzig\c{zwanzig1954high} or Jarynski identities,
\beq
\label{integral1}
\Delta F_{\rm f} = -\ln \int_{-\infty}^{\infty} {\rm d} x\, \e^{-\Delta U(x)} \rho_{\rm s}(x)=-\Delta F_{\rm e}.
\eeq
The value $\Delta F_{\rm f}$ is exactly the negative of the free-energy change \eqq{fec} resulting from the transformation A$\to$B. The total change in free energy is therefore $\Delta F_{\rm tot}=\Delta F_{\rm e}+\Delta F_{\rm f} =0$. This is obvious, because we have started and ended with the same potential, and so the total change in free energy must be zero. 

We then have $\av{W} \geq 0$ by the second law of thermodynamics, and so the measured value of $\av{W} \approx \Delta F_{\rm e}$ is very far from the bound. This discrepancy results from the fact that the final-time change of work is related to the probability of non-erasure, which for large $k$ is very small, while the final-time change in free energy is an exponential average that applies large weight to very rare trajectories that exhibit non-erasure. It is difficult to determine this change of free-energy from numerical evaluation of the Jarzynski identity.  The integral in \eq{integral1} goes as $\sim \int_0^\infty \e^{-k(x-1)^2/2}$, which is dominated by the contribution from the point $x_0=1$. For $k=20$, the likelihood of realizing this value of $x$ is $\rho_{\rm s}(x_0) \approx 10^{-17}$, and so we would need to run more than $10^{17}$ trajectories in order to accurately measure $\Delta F_{\rm tot}=0$ (similar sampling issues have been noted elsewhere\c{berut2013detailed,das2014capturing,buffoni2022spontaneous,vaikuntanathan2009dissipation}).

If we end with the transformation B$\to$B$'$ then we could preserve the bound $\av{W} \geq \Delta F_{\rm s}$ be ensuring that the final-time particle-position distribution $\rho(x(\tf))$ is strictly zero for $x>0$. However, it is difficult to do this for finite-time protocols involving finite potential energy barriers. Conceptually, therefore, it is cleaner to formulate erasure as a compression of configuration space (e.g. the transformation A$\to$B from a double- to a single-well potential) with logical states identified with physical states in a symmetric way. Then the mean work expended in a nonequilibrium erasure process is rigorously bounded as $\av{W} \geq \Delta F_{\rm e}$, which is $\approx \ln 2$ for large enough spring constant\footnote{To model a two-state device we could consider the restoration of the double-well potential to constitute an additional step of the protocol, which could be done with negligible change of work.}. This formulation also makes it simpler to identify the origin of the ``violation'' of the bound  $\av{W} \geq \Delta F_{\rm e}$ for varying-temperature protocols, as we discuss in the following section.

\section{Varying-temperature erasure}
\label{vary}

We now allow the erasure protocol to occur at a non-constant temperature $\beta^{-1}(t)$, with the constraint that $\beta(0)=\beta(\tf)=\beta=1$ (note that $\beta$ with no time argument or subscript label refers to the fixed reciprocal temperature, here chosen to be unity, at the start and end of the protocol). In this case the Jarzynski equality \eq{jarz} is replaced by\c{whitelam2023free}
\beq
\label{jarz_mod}
\av{\e^{-\Omega}}=\e^{-\beta \Delta F}.
\eeq
Here $\Omega \equiv \beta W+\beta Q - \Sigma$, where $Q$ is the heat exchanged with the bath (the energy transferred to the system from the bath), 
\beq
 \label{heat}
 Q = \sum_{i=0}^{N-1} \left[U(\cee_{i+1},x_{i+1})-U(\cee_{i+1},x_i) \right],
 \eeq
 and $-\Sigma$ is the entropy change associated with the trajectory,
 \beq
\Sigma = \sum_{i=0}^{N-1} \beta_{i+1} \left[U(\cee_{i+1},x_{i+1})-U(\cee_{i+1},x_i) \right].
\eeq
In the above expressions the label $i=1,2,\dots,N=\lfloor \tf/\Delta t \rfloor$ refers to the discrete simulation step. The angle brackets in \eq{jarz_mod} denote an average over trajectories that start in equilibrium at temperature $\beta^{-1}$, end at the same temperature (not necessarily in equilibrium), and otherwise involve an arbitrary variation of temperature and other control parameters. The quantity $\Delta F$ is the free-energy change resulting from the protocol, evaluated (as in \eq{jarz}) at temperature $\beta^{-1}$.
\begin{figure}[] 
   \centering
       \ifthenelse{\boolean{retro}}{ \fbox{\includegraphics[width=\linewidth]{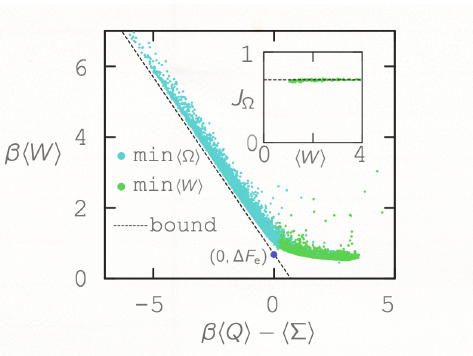}}}
       {\includegraphics[width=\linewidth]{fig_bound}}
   \caption{Averaged values of heat, work, and entropy production for a large number of erasure protocols with non-constant temperature, plotted so that the bound \eq{av_mod} is shown as a black dashed line. The blue dot is the equivalent of the conventional Landauer bound. Inset: Estimator~\eq{est_mod} for a subset of trajectories from the main figure, illustrating the exact result \eq{jarz_mod}. The dotted black line is the free-energy change $\Delta F_{\rm e}$.}
   \label{fig_bound}
\end{figure}

The Jensen inequality applied to \eq{jarz_mod} yields the second law of thermodynamics in the form 
\beq
\label{av_om}
\av{\Omega} \geq \beta \Delta F,
\eeq 
the statement that the total entropy production must be non-negative. This can be written
\beq
\label{av_mod}
\av{W}\geq \Delta F+\beta^{-1}\av{\Sigma}-\av{Q}.
\eeq
Given that the combination $\beta^{-1}\av{\Sigma}-\av{Q}$ can be positive or negative, the statement \eq{av_mod} indicates that if temperature can vary then the work required to do erasure is no longer bounded by $\Delta F=\Delta F_{\rm e}$.

In \f{fig4} we illustrate these relations numerically. We perform the erasure protocol of Figs.~\ref{fig1} and~\ref{fig2} for trajectory length $\tf=10$, but now allow temperature to vary. We set $\beta(t)=\beta'$ for $0<t<\tf$ (with $\beta(0)=\beta(\tf)=1$). We consider a range of $\beta'$ either side of 1.

In \f{fig4}(a) we show the erasure probability resulting from this protocol, as a function of $\beta'$. As $\beta'$ becomes large enough, all of $10^6$ trajectories achieve erasure. 

In \f{fig4}(b) we show that $\av{\Omega}$ is bounded by $\Delta F_{\rm e}$, but the mean work $\av{W}$ is not: for large enough $\beta'$ we observe $\av{W}< \Delta F_{\rm e}$. This is so because the second law applied to varying-temperature processes, \eqq{av_mod}, does not constrain mean work to be greater than the free-energy change for the process. We have ruled out other possibilities for this effect: with our definition of erasure we have $\Delta F_{\rm e} \approx \ln 2$, for large enough $k$, rather than $\Delta F_{\rm e}=0$. We have also shown that $p_{\rm e} \approx 1$, and so the information entropy change upon enacting the process is essentially $\ln 2$.

In \f{fig4}(c) we show that the free-energy estimator
\beq
\label{est_mod}
J_\Omega = -\beta^{-1} \ln \left( N_{\rm traj}^{-1} \sum_{i=1}^{N_{\rm traj}} \e^{-\Omega_i} \right)
\eeq
returns an approximation of the free-energy change $\Delta F_{\rm e}$ for erasure (the estimator \eq{est} applies only to a fixed-temperature protocol). The statistical error in $J_\Omega$ is smallest in the region $\beta' \lesssim 1$, where $\Omega$ and its fluctuations are smallest.

In \f{fig4}(c) we show, for the value $\beta'=0.7$, the validity of the fluctuation relation associated with \eq{jarz_mod}, namely\c{whitelam2023free}
\beq
\label{fluc_mod}
P_{\rm F}(\Omega) \hspace{0.3pt} \e^{-\Omega}=  \e^{-\beta \Delta F} P_{\rm R}(-\Omega).
\eeq
Here $P_{\rm F}(\Omega)$ denotes the probability distribution of $\Omega$ under the erasure protocol, and $P_{\rm R}(-\Omega)$ the distribution of $-\Omega$ under the time-reversed randomization protocol.

In \f{fig_bound} we show that the bound \eq{av_mod} is obeyed for a large number of protocols, for trajectories of length $\tf=10$. Following~\cc{whitelam2023demon} we expressed the time-dependent protocol $(\beta(t),{\bm c}(t))$ using a neural network, and trained the neural network by genetic algorithm to minimize either $\av{\Omega}$ (cyan dots) or $\av{W}$ (green dots). During training, we collected the values of $\av{\Omega}$ and $\av{W}$ generated by each protocol encountered, and plotted them as shown. On this figure the bound \eq{av_mod} is a straight line of slope $-1$ that passes through the point $(0,\Delta F_{\rm e})$. This point, \eqq{av}, is the analog of the constant-temperature Landauer bound for the nonequilibrium formulation of erasure studied here. 

The bound \eq{av_mod} is descriptive (i.e. tight) in the left-hand portion of the figure, and less so in the right-hand portion. By contrast, the relations \eq{jarz_mod} and \eq{fluc_mod} are exact, and provide more detailed statements than the bound \eq{av_mod} about the relationship between work and entropy production for varying-temperature protocols. In the inset of \f{fig_bound} we show that~\eqq{jarz_mod} holds for these protocols\footnote{Note that $\Omega = \beta W$ when temperature is fixed, in which case \eq{jarz_mod} and \eq{fluc_mod} reduce to the standard relations \eq{jarz} and \eq{fluc}, respectively.}.

We note that the bound \eq{av_mod} is equivalent to that given in~\cc{wolpert2023stochastic}: from the first law of thermodynamics we have $W+Q=\Delta U$, the change of internal energy of the system, and so \eq{av_mod} can be written $\av{\Delta U}-\Delta F \geq\beta^{-1}\av{\Sigma}$. Noting that $\Delta F= \av{\Delta U} -\beta^{-1} \av{\Delta S}$, where $\Delta S$ is the change in entropy of the system, we can write \eq{av_mod} as $\av{\Sigma} \leq \av{\Delta S}$. This is the bound resulting from the non-negativity of Eq. (1) of~\cc{wolpert2023stochastic} (noting that, in that paper, $Q$ is heat transferred from the system to the bath, and so is the negative of the heat $Q$ considered here).

\begin{figure}[] 
   \centering
       \ifthenelse{\boolean{retro}}{  \fbox{\includegraphics[width=\linewidth]{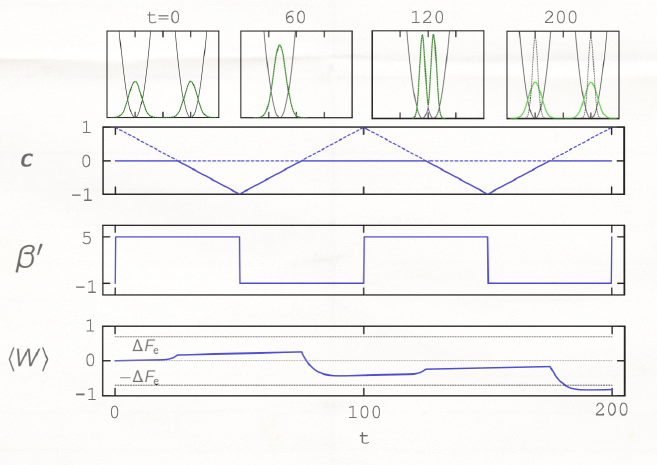}}}
       {\includegraphics[width=\linewidth]{fig_engine}}
   \caption{Two cycles of erasure and randomization, running the erasure protocol of \f{fig1}(a) and its time reverse at different temperatures. Under this cycle, work can be extracted. Averages are calculated over $10^5$ independent trajectories. Snapshots at the top have the same format as \f{fig1}(b).}
   \label{fig_engine}
\end{figure}

One consequence of the inequality \eq{av_mod} is that a varying-temperature erasure-randomization cycle can be run as a work-producing engine. As shown in \f{fig3}(b), erasure carried out at a suitably large value of $\beta'$ can be done with less work than $\Delta F_{\rm e}$. If we randomize the memory at reciprocal temperature $\beta'=1$, we can extract work up to an amount $\Delta F_{\rm e}$. We should therefore be able to extract work from the erasure-randomization cycle, if performed slowly enough. In \f{fig_engine} we confirm this prediction numerically. We perform a varying-temperature cycle of erasure, randomization, erasure, and randomization, with each component of the cycle lasting for time $t=50$. As shown in the bottom panel, work is extracted.

\section{Conclusions}
\label{conc}

We have investigated bit erasure done at varying temperature, to which Landauer's result does not apply. We have formulated erasure as a stochastic nonequilibrium process involving a compression of configuration space. Erasure starts and ends at temperature $T$, but temperature can otherwise vary with time in an arbitrary way. 
Defined in this way, erasure is described by a set of nonequilibrium fluctuation relations that place limits on the efficiency of the process and its reverse, bit randomization. Under a varying-temperature protocol, erasure and randomization can be operated as a work-producing cycle.

\section{Acknowledgments} This work was performed at the Molecular Foundry at Lawrence Berkeley National Laboratory, \ifthenelse{\boolean{retro}}{with costs defrayed by}{supported by} the Office of Basic Energy Sciences of the U.S. Department of Energy under Contract No. DE-AC02--05CH11231.


\begin{thebibliography}{30}%
\makeatletter
\providecommand \@ifxundefined [1]{%
 \@ifx{#1\undefined}
}%
\providecommand \@ifnum [1]{%
 \ifnum #1\expandafter \@firstoftwo
 \else \expandafter \@secondoftwo
 \fi
}%
\providecommand \@ifx [1]{%
 \ifx #1\expandafter \@firstoftwo
 \else \expandafter \@secondoftwo
 \fi
}%
\providecommand \natexlab [1]{#1}%
\providecommand \enquote  [1]{``#1''}%
\providecommand \bibnamefont  [1]{#1}%
\providecommand \bibfnamefont [1]{#1}%
\providecommand \citenamefont [1]{#1}%
\providecommand \href@noop [0]{\@secondoftwo}%
\providecommand \href [0]{\begingroup \@sanitize@url \@href}%
\providecommand \@href[1]{\@@startlink{#1}\@@href}%
\providecommand \@@href[1]{\endgroup#1\@@endlink}%
\providecommand \@sanitize@url [0]{\catcode `\\12\catcode `\$12\catcode
  `\&12\catcode `\#12\catcode `\^12\catcode `\_12\catcode `\%12\relax}%
\providecommand \@@startlink[1]{}%
\providecommand \@@endlink[0]{}%
\providecommand \url  [0]{\begingroup\@sanitize@url \@url }%
\providecommand \@url [1]{\endgroup\@href {#1}{\urlprefix }}%
\providecommand \urlprefix  [0]{URL }%
\providecommand \Eprint [0]{\href }%
\providecommand \doibase [0]{http://dx.doi.org/}%
\providecommand \selectlanguage [0]{\@gobble}%
\providecommand \bibinfo  [0]{\@secondoftwo}%
\providecommand \bibfield  [0]{\@secondoftwo}%
\providecommand \translation [1]{[#1]}%
\providecommand \BibitemOpen [0]{}%
\providecommand \bibitemStop [0]{}%
\providecommand \bibitemNoStop [0]{.\EOS\space}%
\providecommand \EOS [0]{\spacefactor3000\relax}%
\providecommand \BibitemShut  [1]{\csname bibitem#1\endcsname}%
\let\auto@bib@innerbib\@empty
\bibitem [{\citenamefont {Landauer}(1961)}]{landauer1961irreversibility}%
  \BibitemOpen
  \bibfield  {author} {\bibinfo {author} {\bibfnamefont {R.}~\bibnamefont
  {Landauer}},\ }\href@noop {} {\bibfield  {journal} {\bibinfo  {journal} {IBM
  journal of research and development}\ }\textbf {\bibinfo {volume} {5}},\
  \bibinfo {pages} {183} (\bibinfo {year} {1961})}\BibitemShut {NoStop}%
\bibitem [{\citenamefont {Sagawa}(2014)}]{sagawa2014thermodynamic}%
  \BibitemOpen
  \bibfield  {author} {\bibinfo {author} {\bibfnamefont {T.}~\bibnamefont
  {Sagawa}},\ }\href@noop {} {\bibfield  {journal} {\bibinfo  {journal}
  {Journal of Statistical Mechanics: Theory and Experiment}\ }\textbf {\bibinfo
  {volume} {2014}},\ \bibinfo {pages} {P03025} (\bibinfo {year}
  {2014})}\BibitemShut {NoStop}%
\bibitem [{\citenamefont {Bennett}(1982)}]{bennett1982thermodynamics}%
  \BibitemOpen
  \bibfield  {author} {\bibinfo {author} {\bibfnamefont {C.~H.}\ \bibnamefont
  {Bennett}},\ }\href@noop {} {\bibfield  {journal} {\bibinfo  {journal}
  {International Journal of Theoretical Physics}\ }\textbf {\bibinfo {volume}
  {21}},\ \bibinfo {pages} {905} (\bibinfo {year} {1982})}\BibitemShut
  {NoStop}%
\bibitem [{\citenamefont {Bennett}\ and\ \citenamefont
  {{L}andauer}(1985)}]{bennett1985fundamental}%
  \BibitemOpen
  \bibfield  {author} {\bibinfo {author} {\bibfnamefont {C.~H.}\ \bibnamefont
  {Bennett}}\ and\ \bibinfo {author} {\bibfnamefont {R.}~\bibnamefont
  {{L}andauer}},\ }\href@noop {} {\bibfield  {journal} {\bibinfo  {journal}
  {Scientific American}\ }\textbf {\bibinfo {volume} {253}},\ \bibinfo {pages}
  {48} (\bibinfo {year} {1985})}\BibitemShut {NoStop}%
\bibitem [{\citenamefont {Piechocinska}(2000)}]{piechocinska2000information}%
  \BibitemOpen
  \bibfield  {author} {\bibinfo {author} {\bibfnamefont {B.}~\bibnamefont
  {Piechocinska}},\ }\href@noop {} {\bibfield  {journal} {\bibinfo  {journal}
  {Physical Review A}\ }\textbf {\bibinfo {volume} {61}},\ \bibinfo {pages}
  {062314} (\bibinfo {year} {2000})}\BibitemShut {NoStop}%
\bibitem [{\citenamefont {Dillenschneider}\ and\ \citenamefont
  {Lutz}(2009)}]{dillenschneider2009memory}%
  \BibitemOpen
  \bibfield  {author} {\bibinfo {author} {\bibfnamefont {R.}~\bibnamefont
  {Dillenschneider}}\ and\ \bibinfo {author} {\bibfnamefont {E.}~\bibnamefont
  {Lutz}},\ }\href@noop {} {\bibfield  {journal} {\bibinfo  {journal} {Physical
  Review Letters}\ }\textbf {\bibinfo {volume} {102}},\ \bibinfo {pages}
  {210601} (\bibinfo {year} {2009})}\BibitemShut {NoStop}%
\bibitem [{\citenamefont {Jun}\ \emph {et~al.}(2014)\citenamefont {Jun},
  \citenamefont {Gavrilov},\ and\ \citenamefont {Bechhoefer}}]{jun2014high}%
  \BibitemOpen
  \bibfield  {author} {\bibinfo {author} {\bibfnamefont {Y.}~\bibnamefont
  {Jun}}, \bibinfo {author} {\bibfnamefont {M.}~\bibnamefont {Gavrilov}}, \
  and\ \bibinfo {author} {\bibfnamefont {J.}~\bibnamefont {Bechhoefer}},\
  }\href@noop {} {\bibfield  {journal} {\bibinfo  {journal} {Physical Review
  Letters}\ }\textbf {\bibinfo {volume} {113}},\ \bibinfo {pages} {190601}
  (\bibinfo {year} {2014})}\BibitemShut {NoStop}%
\bibitem [{\citenamefont {B{\'e}rut}\ \emph {et~al.}(2012)\citenamefont
  {B{\'e}rut}, \citenamefont {Arakelyan}, \citenamefont {Petrosyan},
  \citenamefont {Ciliberto}, \citenamefont {Dillenschneider},\ and\
  \citenamefont {Lutz}}]{berut2012experimental}%
  \BibitemOpen
  \bibfield  {author} {\bibinfo {author} {\bibfnamefont {A.}~\bibnamefont
  {B{\'e}rut}}, \bibinfo {author} {\bibfnamefont {A.}~\bibnamefont
  {Arakelyan}}, \bibinfo {author} {\bibfnamefont {A.}~\bibnamefont
  {Petrosyan}}, \bibinfo {author} {\bibfnamefont {S.}~\bibnamefont
  {Ciliberto}}, \bibinfo {author} {\bibfnamefont {R.}~\bibnamefont
  {Dillenschneider}}, \ and\ \bibinfo {author} {\bibfnamefont {E.}~\bibnamefont
  {Lutz}},\ }\href@noop {} {\bibfield  {journal} {\bibinfo  {journal} {Nature}\
  }\textbf {\bibinfo {volume} {483}},\ \bibinfo {pages} {187} (\bibinfo {year}
  {2012})}\BibitemShut {NoStop}%
\bibitem [{\citenamefont {Dago}\ \emph {et~al.}(2021)\citenamefont {Dago},
  \citenamefont {Pereda}, \citenamefont {Barros}, \citenamefont {Ciliberto},\
  and\ \citenamefont {Bellon}}]{dago2021information}%
  \BibitemOpen
  \bibfield  {author} {\bibinfo {author} {\bibfnamefont {S.}~\bibnamefont
  {Dago}}, \bibinfo {author} {\bibfnamefont {J.}~\bibnamefont {Pereda}},
  \bibinfo {author} {\bibfnamefont {N.}~\bibnamefont {Barros}}, \bibinfo
  {author} {\bibfnamefont {S.}~\bibnamefont {Ciliberto}}, \ and\ \bibinfo
  {author} {\bibfnamefont {L.}~\bibnamefont {Bellon}},\ }\href@noop {}
  {\bibfield  {journal} {\bibinfo  {journal} {Physical Review Letters}\
  }\textbf {\bibinfo {volume} {126}},\ \bibinfo {pages} {170601} (\bibinfo
  {year} {2021})}\BibitemShut {NoStop}%
\bibitem [{\citenamefont {Hong}\ \emph {et~al.}(2016)\citenamefont {Hong},
  \citenamefont {Lambson}, \citenamefont {Dhuey},\ and\ \citenamefont
  {Bokor}}]{hong2016experimental}%
  \BibitemOpen
  \bibfield  {author} {\bibinfo {author} {\bibfnamefont {J.}~\bibnamefont
  {Hong}}, \bibinfo {author} {\bibfnamefont {B.}~\bibnamefont {Lambson}},
  \bibinfo {author} {\bibfnamefont {S.}~\bibnamefont {Dhuey}}, \ and\ \bibinfo
  {author} {\bibfnamefont {J.}~\bibnamefont {Bokor}},\ }\href@noop {}
  {\bibfield  {journal} {\bibinfo  {journal} {Science {A}dvances}\ }\textbf
  {\bibinfo {volume} {2}},\ \bibinfo {pages} {e1501492} (\bibinfo {year}
  {2016})}\BibitemShut {NoStop}%
\bibitem [{\citenamefont {Frank}(2018)}]{frank2018physical}%
  \BibitemOpen
  \bibfield  {author} {\bibinfo {author} {\bibfnamefont {M.~P.}\ \bibnamefont
  {Frank}},\ }in\ \href@noop {} {\emph {\bibinfo {booktitle} {International
  Conference on Reversible Computation}}}\ (\bibinfo {organization}
  {Springer},\ \bibinfo {year} {2018})\ pp.\ \bibinfo {pages}
  {3--33}\BibitemShut {NoStop}%
\bibitem [{\citenamefont {Frank}(2005)}]{frank2005approaching}%
  \BibitemOpen
  \bibfield  {author} {\bibinfo {author} {\bibfnamefont {M.~P.}\ \bibnamefont
  {Frank}},\ }in\ \href@noop {} {\emph {\bibinfo {booktitle} {35th
  International Symposium on Multiple-Valued Logic (ISMVL'05)}}}\ (\bibinfo
  {organization} {IEEE},\ \bibinfo {year} {2005})\ pp.\ \bibinfo {pages}
  {168--185}\BibitemShut {NoStop}%
\bibitem [{\citenamefont {Still}\ \emph {et~al.}(2012)\citenamefont {Still},
  \citenamefont {Sivak}, \citenamefont {Bell},\ and\ \citenamefont
  {Crooks}}]{still2012thermodynamics}%
  \BibitemOpen
  \bibfield  {author} {\bibinfo {author} {\bibfnamefont {S.}~\bibnamefont
  {Still}}, \bibinfo {author} {\bibfnamefont {D.~A.}\ \bibnamefont {Sivak}},
  \bibinfo {author} {\bibfnamefont {A.~J.}\ \bibnamefont {Bell}}, \ and\
  \bibinfo {author} {\bibfnamefont {G.~E.}\ \bibnamefont {Crooks}},\
  }\href@noop {} {\bibfield  {journal} {\bibinfo  {journal} {Physical review
  letters}\ }\textbf {\bibinfo {volume} {109}},\ \bibinfo {pages} {120604}
  (\bibinfo {year} {2012})}\BibitemShut {NoStop}%
\bibitem [{\citenamefont {Maroney}(2005)}]{maroney2005absence}%
  \BibitemOpen
  \bibfield  {author} {\bibinfo {author} {\bibfnamefont {O.~J.}\ \bibnamefont
  {Maroney}},\ }\href@noop {} {\bibfield  {journal} {\bibinfo  {journal}
  {Studies in History and Philosophy of Science Part B: Studies in History and
  Philosophy of Modern Physics}\ }\textbf {\bibinfo {volume} {36}},\ \bibinfo
  {pages} {355} (\bibinfo {year} {2005})}\BibitemShut {NoStop}%
\bibitem [{\citenamefont {Maroney}(2009)}]{maroney2009generalizing}%
  \BibitemOpen
  \bibfield  {author} {\bibinfo {author} {\bibfnamefont {O.~J.}\ \bibnamefont
  {Maroney}},\ }\href@noop {} {\bibfield  {journal} {\bibinfo  {journal}
  {Physical Review E}\ }\textbf {\bibinfo {volume} {79}},\ \bibinfo {pages}
  {031105} (\bibinfo {year} {2009})}\BibitemShut {NoStop}%
\bibitem [{\citenamefont {Seet}\ \emph {et~al.}(2023)\citenamefont {Seet},
  \citenamefont {Ouldridge},\ and\ \citenamefont {Doye}}]{seet2023simulation}%
  \BibitemOpen
  \bibfield  {author} {\bibinfo {author} {\bibfnamefont {I.}~\bibnamefont
  {Seet}}, \bibinfo {author} {\bibfnamefont {T.~E.}\ \bibnamefont {Ouldridge}},
  \ and\ \bibinfo {author} {\bibfnamefont {J.~P.}\ \bibnamefont {Doye}},\
  }\href@noop {} {\bibfield  {journal} {\bibinfo  {journal} {Physical Review
  E}\ }\textbf {\bibinfo {volume} {107}},\ \bibinfo {pages} {024134} (\bibinfo
  {year} {2023})}\BibitemShut {NoStop}%
\bibitem [{\citenamefont {Wolpert}\ \emph {et~al.}(2023)\citenamefont
  {Wolpert}, \citenamefont {Korbel}, \citenamefont {Lynn}, \citenamefont
  {Tasnim}, \citenamefont {Grochow}, \citenamefont {Kardes}, \citenamefont
  {Aimone}, \citenamefont {Balasubramanian}, \citenamefont {de~Giuli},
  \citenamefont {Doty} \emph {et~al.}}]{wolpert2023stochastic}%
  \BibitemOpen
  \bibfield  {author} {\bibinfo {author} {\bibfnamefont {D.}~\bibnamefont
  {Wolpert}}, \bibinfo {author} {\bibfnamefont {J.}~\bibnamefont {Korbel}},
  \bibinfo {author} {\bibfnamefont {C.}~\bibnamefont {Lynn}}, \bibinfo {author}
  {\bibfnamefont {F.}~\bibnamefont {Tasnim}}, \bibinfo {author} {\bibfnamefont
  {J.}~\bibnamefont {Grochow}}, \bibinfo {author} {\bibfnamefont
  {G.}~\bibnamefont {Kardes}}, \bibinfo {author} {\bibfnamefont
  {J.}~\bibnamefont {Aimone}}, \bibinfo {author} {\bibfnamefont
  {V.}~\bibnamefont {Balasubramanian}}, \bibinfo {author} {\bibfnamefont
  {E.}~\bibnamefont {de~Giuli}}, \bibinfo {author} {\bibfnamefont
  {D.}~\bibnamefont {Doty}},  \emph {et~al.},\ }\href@noop {} {\bibfield
  {journal} {\bibinfo  {journal} {arXiv preprint arXiv:2311.17166}\ } (\bibinfo
  {year} {2023})}\BibitemShut {NoStop}%
\bibitem [{\citenamefont {Gingrich}\ \emph {et~al.}(2016)\citenamefont
  {Gingrich}, \citenamefont {Rotskoff}, \citenamefont {Crooks},\ and\
  \citenamefont {Geissler}}]{gingrich2016near}%
  \BibitemOpen
  \bibfield  {author} {\bibinfo {author} {\bibfnamefont {T.~R.}\ \bibnamefont
  {Gingrich}}, \bibinfo {author} {\bibfnamefont {G.~M.}\ \bibnamefont
  {Rotskoff}}, \bibinfo {author} {\bibfnamefont {G.~E.}\ \bibnamefont
  {Crooks}}, \ and\ \bibinfo {author} {\bibfnamefont {P.~L.}\ \bibnamefont
  {Geissler}},\ }\href@noop {} {\bibfield  {journal} {\bibinfo  {journal}
  {Proceedings of the National Academy of Sciences}\ }\textbf {\bibinfo
  {volume} {113}},\ \bibinfo {pages} {10263} (\bibinfo {year}
  {2016})}\BibitemShut {NoStop}%
\bibitem [{\citenamefont {Rotskoff}\ and\ \citenamefont
  {Crooks}(2015)}]{rotskoff2015optimal}%
  \BibitemOpen
  \bibfield  {author} {\bibinfo {author} {\bibfnamefont {G.~M.}\ \bibnamefont
  {Rotskoff}}\ and\ \bibinfo {author} {\bibfnamefont {G.~E.}\ \bibnamefont
  {Crooks}},\ }\href@noop {} {\bibfield  {journal} {\bibinfo  {journal}
  {Physical Review E}\ }\textbf {\bibinfo {volume} {92}},\ \bibinfo {pages}
  {060102} (\bibinfo {year} {2015})}\BibitemShut {NoStop}%
\bibitem [{\citenamefont {Engel}\ \emph {et~al.}(2023)\citenamefont {Engel},
  \citenamefont {Smith},\ and\ \citenamefont {Brenner}}]{engel2023optimal}%
  \BibitemOpen
  \bibfield  {author} {\bibinfo {author} {\bibfnamefont {M.~C.}\ \bibnamefont
  {Engel}}, \bibinfo {author} {\bibfnamefont {J.~A.}\ \bibnamefont {Smith}}, \
  and\ \bibinfo {author} {\bibfnamefont {M.~P.}\ \bibnamefont {Brenner}},\
  }\href@noop {} {\bibfield  {journal} {\bibinfo  {journal} {Physical Review
  X}\ }\textbf {\bibinfo {volume} {13}},\ \bibinfo {pages} {041032} (\bibinfo
  {year} {2023})}\BibitemShut {NoStop}%
\bibitem [{\citenamefont {Crooks}(1999)}]{crooks1999entropy}%
  \BibitemOpen
  \bibfield  {author} {\bibinfo {author} {\bibfnamefont {G.~E.}\ \bibnamefont
  {Crooks}},\ }\href@noop {} {\bibfield  {journal} {\bibinfo  {journal}
  {Physical Review E}\ }\textbf {\bibinfo {volume} {60}},\ \bibinfo {pages}
  {2721} (\bibinfo {year} {1999})}\BibitemShut {NoStop}%
\bibitem [{\citenamefont {Jarzynski}(1997)}]{jarzynski1997nonequilibrium}%
  \BibitemOpen
  \bibfield  {author} {\bibinfo {author} {\bibfnamefont {C.}~\bibnamefont
  {Jarzynski}},\ }\href@noop {} {\bibfield  {journal} {\bibinfo  {journal}
  {Physical Review Letters}\ }\textbf {\bibinfo {volume} {78}},\ \bibinfo
  {pages} {2690} (\bibinfo {year} {1997})}\BibitemShut {NoStop}%
\bibitem [{\citenamefont {Das}(2014)}]{das2014capturing}%
  \BibitemOpen
  \bibfield  {author} {\bibinfo {author} {\bibfnamefont {M.}~\bibnamefont
  {Das}},\ }\href@noop {} {\bibfield  {journal} {\bibinfo  {journal} {Physical
  Review E}\ }\textbf {\bibinfo {volume} {90}},\ \bibinfo {pages} {062120}
  (\bibinfo {year} {2014})}\BibitemShut {NoStop}%
\bibitem [{\citenamefont {Whitelam}(2023{\natexlab{a}})}]{whitelam2023free}%
  \BibitemOpen
  \bibfield  {author} {\bibinfo {author} {\bibfnamefont {S.}~\bibnamefont
  {Whitelam}},\ }\href@noop {} {\bibfield  {journal} {\bibinfo  {journal}
  {arXiv preprint arXiv:2311.06997}\ } (\bibinfo {year}
  {2023}{\natexlab{a}})}\BibitemShut {NoStop}%
\bibitem [{\citenamefont {Dago}\ \emph {et~al.}(2022)\citenamefont {Dago},
  \citenamefont {Pereda}, \citenamefont {Ciliberto},\ and\ \citenamefont
  {Bellon}}]{dago2022virtual}%
  \BibitemOpen
  \bibfield  {author} {\bibinfo {author} {\bibfnamefont {S.}~\bibnamefont
  {Dago}}, \bibinfo {author} {\bibfnamefont {J.}~\bibnamefont {Pereda}},
  \bibinfo {author} {\bibfnamefont {S.}~\bibnamefont {Ciliberto}}, \ and\
  \bibinfo {author} {\bibfnamefont {L.}~\bibnamefont {Bellon}},\ }\href@noop {}
  {\bibfield  {journal} {\bibinfo  {journal} {Journal of Statistical Mechanics:
  Theory and Experiment}\ }\textbf {\bibinfo {volume} {2022}},\ \bibinfo
  {pages} {053209} (\bibinfo {year} {2022})}\BibitemShut {NoStop}%
\bibitem [{\citenamefont {Zwanzig}(1954)}]{zwanzig1954high}%
  \BibitemOpen
  \bibfield  {author} {\bibinfo {author} {\bibfnamefont {R.~W.}\ \bibnamefont
  {Zwanzig}},\ }\href@noop {} {\bibfield  {journal} {\bibinfo  {journal} {The
  Journal of Chemical Physics}\ }\textbf {\bibinfo {volume} {22}},\ \bibinfo
  {pages} {1420} (\bibinfo {year} {1954})}\BibitemShut {NoStop}%
\bibitem [{\citenamefont {B{\'e}rut}\ \emph {et~al.}(2013)\citenamefont
  {B{\'e}rut}, \citenamefont {Petrosyan},\ and\ \citenamefont
  {Ciliberto}}]{berut2013detailed}%
  \BibitemOpen
  \bibfield  {author} {\bibinfo {author} {\bibfnamefont {A.}~\bibnamefont
  {B{\'e}rut}}, \bibinfo {author} {\bibfnamefont {A.}~\bibnamefont
  {Petrosyan}}, \ and\ \bibinfo {author} {\bibfnamefont {S.}~\bibnamefont
  {Ciliberto}},\ }\href@noop {} {\bibfield  {journal} {\bibinfo  {journal}
  {Europhysics Letters}\ }\textbf {\bibinfo {volume} {103}},\ \bibinfo {pages}
  {60002} (\bibinfo {year} {2013})}\BibitemShut {NoStop}%
\bibitem [{\citenamefont {Buffoni}\ and\ \citenamefont
  {Campisi}(2022)}]{buffoni2022spontaneous}%
  \BibitemOpen
  \bibfield  {author} {\bibinfo {author} {\bibfnamefont {L.}~\bibnamefont
  {Buffoni}}\ and\ \bibinfo {author} {\bibfnamefont {M.}~\bibnamefont
  {Campisi}},\ }\href@noop {} {\bibfield  {journal} {\bibinfo  {journal}
  {Journal of Statistical Physics}\ }\textbf {\bibinfo {volume} {186}},\
  \bibinfo {pages} {31} (\bibinfo {year} {2022})}\BibitemShut {NoStop}%
\bibitem [{\citenamefont {Vaikuntanathan}\ and\ \citenamefont
  {Jarzynski}(2009)}]{vaikuntanathan2009dissipation}%
  \BibitemOpen
  \bibfield  {author} {\bibinfo {author} {\bibfnamefont {S.}~\bibnamefont
  {Vaikuntanathan}}\ and\ \bibinfo {author} {\bibfnamefont {C.}~\bibnamefont
  {Jarzynski}},\ }\href@noop {} {\bibfield  {journal} {\bibinfo  {journal}
  {Europhysics Letters}\ }\textbf {\bibinfo {volume} {87}},\ \bibinfo {pages}
  {60005} (\bibinfo {year} {2009})}\BibitemShut {NoStop}%
\bibitem [{\citenamefont {Whitelam}(2023{\natexlab{b}})}]{whitelam2023demon}%
  \BibitemOpen
  \bibfield  {author} {\bibinfo {author} {\bibfnamefont {S.}~\bibnamefont
  {Whitelam}},\ }\href@noop {} {\bibfield  {journal} {\bibinfo  {journal}
  {Physical Review X}\ }\textbf {\bibinfo {volume} {13}},\ \bibinfo {pages}
  {021005} (\bibinfo {year} {2023}{\natexlab{b}})}\BibitemShut {NoStop}%
\end{thebibliography}

%

\end{document}